\documentclass[useAMS,usenatbib]{mn2e}

\usepackage{graphicx}
\usepackage{txfonts}
\title[HerMES: SDP Maps]{HerMES: SPIRE Science Demonstration Phase Maps\thanks{ {\sl Herschel\/} is an ESA space observatory with science instruments 
provided by European-led Principal Investigator consortia and with important participation from NASA.}\thanks{hermes.sussex.ac.uk}\thanks{hedam.oamp.fr}}
\author[L.~Levenson et al.]
{\parbox{\textwidth}{\raggedright L.~Levenson,$^{1,2}$\thanks{E-mail: \texttt{levenson@caltech.edu}}
G.~Marsden,$^{3}$
M.~Zemcov,$^{1,2}$
A.~Amblard,$^{4}$
A.~Blain,$^{1}$
J.~Bock,$^{1,2}$
E.~Chapin,$^{3}$
A.~Conley,$^{5}$
A.~Cooray,$^{4,1}$
C.D.~Dowell,$^{1,2}$
T.P.~Ellsworth-Bowers,$^{5}$
A.~Franceschini,$^{6}$
J.~Glenn,$^{5}$
M.~Griffin,$^{7}$
M.~Halpern,$^{3}$
H.T.~Nguyen,$^{2,1}$
S.J.~Oliver,$^{8}$
M.J.~Page,$^{9}$
A.~Papageorgiou,$^{7}$
I.~P{\'e}rez-Fournon,$^{10,11}$
M.~Pohlen,$^{7}$
N.~Rangwala,$^{5}$
M.~Rowan-Robinson,$^{12}$
B.~Schulz,$^{1,13}$
Douglas~Scott,$^{3}$
P.~Serra,$^{4}$
D.L.~Shupe,$^{1,13}$
E.~Valiante,$^{3}$
J.D.~Vieira,$^{1}$
L.~Vigroux,$^{14}$
D.~Wiebe,$^{3}$
G.~Wright$^{15}$ and
C.K.~Xu$^{1,13}$}\vspace{0.4cm}\\
\parbox{\textwidth}{\raggedright $^{1}$California Institute of Technology, 1200 E. California Blvd., Pasadena, CA 91125, USA\\
$^{2}$Jet Propulsion Laboratory, 4800 Oak Grove Drive, Pasadena, CA 91109, USA\\
$^{3}$Department of Physics \& Astronomy, University of British Columbia, 6224 Agricultural Road, Vancouver, BC V6T~1Z1, Canada\\
$^{4}$Dept. of Physics \& Astronomy, University of California, Irvine, CA 92697, USA\\
$^{5}$Dept. of Astrophysical and Planetary Sciences, CASA 389-UCB, University of Colorado, Boulder, CO 80309, USA\\
$^{6}$Dipartimento di Astronomia, Universit\`{a} di Padova, vicolo Osservatorio, 3, 35122 Padova, Italy\\
$^{7}$Cardiff School of Physics and Astronomy, Cardiff University, Queens Buildings, The Parade, Cardiff CF24 3AA, UK\\
$^{8}$Astronomy Centre, Dept. of Physics \& Astronomy, University of Sussex, Brighton BN1 9QH, UK\\
$^{9}$Mullard Space Science Laboratory, University College London, Holmbury St. Mary, Dorking, Surrey RH5 6NT, UK\\
$^{10}$Instituto de Astrof{\'\i}sica de Canarias (IAC), E-38200 La Laguna, Tenerife, Spain\\
$^{11}$Departamento de Astrof{\'\i}sica, Universidad de La Laguna (ULL), E-38205 La Laguna, Tenerife, Spain\\
$^{12}$Astrophysics Group, Imperial College London, Blackett Laboratory, Prince Consort Road, London SW7 2AZ, UK\\
$^{13}$Infrared Processing and Analysis Center, MS 100-22, California Institute of Technology, JPL, Pasadena, CA 91125, USA\\
$^{14}$Institut d'Astrophysique de Paris, UMR 7095, CNRS, UPMC Univ. Paris 06, 98bis boulevard Arago, F-75014 Paris, France\\
$^{15}$UK Astronomy Technology Centre, Royal Observatory, Blackford Hill, Edinburgh EH9 3HJ, UK}}
\begin{document}

\date{MNRAS Accepted -- 27 September, 2010}

\pagerange{\pageref{firstpage}--\pageref{lastpage}} \pubyear{2002}

\maketitle

\label{firstpage}

\begin{abstract}
We describe the production and verification of sky maps of the five SPIRE fields observed as part of the {\sl Herschel\/} Multi-tiered Extragalactic Survey (HerMES) during the Science Demonstration Phase (SDP) of the {\sl Herschel\/} mission. We have implemented an iterative map-making algorithm (SHIM; The SPIRE-HerMES Iterative Mapper) to produce high fidelity maps that preserve extended diffuse emission on the sky while exploiting the repeated observations of the same region of the sky with many detectors in multiple scan directions to minimize residual instrument noise. We specify here the SHIM algorithm and  outline the various tests that were performed to determine and characterize the quality of the maps and verify that the astrometry, point source flux and power on all relevant angular scales meets the needs of the HerMES science goals.  These include multiple jackknife tests, determination of the map transfer function and detailed examination of the power spectra of both sky and jackknife maps. The map transfer function is approximately unity on scales from one arcminute to one degree.  Final maps (v1.0), including multiple jackknives, as well as the SHIM pipeline, have been used by the HerMES team for the production of SDP papers. \end{abstract}


\section{Introduction}

The Spectral and Photometric Imaging Receiver 
(SPIRE; \citealt{griffin}) on board the {\sl Herschel\/} Space Observatory
\citep{pilbrat} has opened a new window on the Universe at
far-infrared/submillimetre (FIR/submm) wavelengths.  The spectacular sensitivity of SPIRE
detectors combined with {\sl Herschel\/}'s 3.5\,m aperture, currently the
largest telescope in space, allows astronomers to observe the FIR/submm
Universe with unprecedented efficiency. During the Science Demonstration Phase (SDP) of the {\sl Herschel\/} mission,
five fields of various size and depth were observed as part of the {\sl Herschel\/}
Multi-tiered Extragalactic Survey \citep[HeMES,][]{oliver2}.

The SPIRE photometer focal plane arrays (FPAs) consist of 139, 88 and 43 feedhorn-coupled spider-web bolometers which measure temperature changes resulting from the absorption of incident radiation in three bands centered on 250, 350 and 500\,\micron\ respectively.   The {\sl Herschel\/} Interactive Processing Environment ({\sc hipe\/}\footnote{http://herschel.esac.esa.int/{\sc hipe\/}\_download.shtml}, \citealt{ott}) data processing pipeline, which converts raw bolometer data to physical units and removes artefacts like cosmic rays and  large-scale signal variations due to the  temperature changes in the FPAs are removed is described in detail in \citet{dowell}.   

This pipeline is broken into several stages, including the conversion of raw telescope data into engineering units (i.e. Volts), the conversion of voltage into flux density units, corrections for electrical, optical, thermal and astrophysical artefacts and, finally, mapping of the time-ordered data into sky images.  Details of the {\sc hipe\/} settings used for the current processing are given in Section~\ref{timestreams}.  {\sc hipe\/} works on detector timelines which consist of the detector samples for a single scan, at constant scan speed, across the map area.  Currently, the default map-maker in {\sc hipe\/} removes a median from each detector timeline and then simply bins the measured fluxes into map pixels.  For the majority of the HerMES Science Demonstration Phase (SDP) science goals, particularly those studies that focus on the detection and characterization of individual sources, the {\sc hipe\/} `naive' maps meet all the requirements.  However, science goals that depend on the statistical power in the map to measure large-scale structure and constrain source count models, are particularly sensitive to artefacts and/or residual power (or lack thereof) in sky maps due to the processing.  

The SPIRE FPAs are held at $\approx$ 0.3K and the temperature fluctuations due to sources on the sky are generally much smaller than slow temperature drifts in the focal plane structure itself. For this reason, a single, raw scan of the sky is dominated by low frequency noise, correlated among detectors, and commonly referred to as $1/f $ noise.  The SPIRE instrument tracks these drifts using both thermistors on the focal plane and dark bolometers that are not exposed to the sky.  These additional data allow removal of the $1/f$ noise by the {\sc hipe\/} pipeline.  The drifts are removed so completely by the standard pipeline that, as stated above, for studies of individual sources, there is no appreciable gain in achieved sensitivity in going from a simple binning mapper to a more sophisticated method.  For statistical studies, however, any subtle phase difference in thermistor/bolometer response, an uncoupled bolometer drift, or an error in interpreting the thermistor coupling to the bolometer signal can introduce or leave false signal in the maps that, while difficult to detect, can affect science results.  Additionally, the process of binning time-ordered data into sky maps must be optimized to accurately reproduce optical power on all relevant angular scales and must be characterized for potential map making artefacts that can bias statistical interpretations of map data.  

This paper describes the production and characterization of a set of high-fidelity maps, now in use by the  HerMES team, produced with a mapping algorithm designed to mitigate these potential artefacts, should they exist in the post-HIPE data.  

In selecting a mapping algorithm for the HerMES SDP maps, it was anticipated, based on ground-based and balloon-borne FIR/submm/microwave experiments, that maximum likelihood map-makers, such as MADmap \citep{madmap} or SANEPIC \citep{pat09} would be needed.  While MADmap has been implemented in {\sc hipe\/}, it is not optimal for SPIRE data due to the large number of detectors with correlated noise.  SANEPIC was designed for the data from BLAST, \citep[the Balloon Borne Large Area Submillimeter Telescope,][]{pat09} which flew a prototype of the SPIRE instrument on a stratospheric balloon and is an optimal map-maker made to specifically handle correlated noise among many detectors.  Ultimately a SANEPIC-like optimal solution will be used to map the SPIRE data.  However, the environment at the second Lagrange point (L2) of the Sun-Earth system, 1.5 million km from Earth, is far more stable than any environment here on Earth, as well as having no atmosphere.  Thus, it was determined early in the {\sl Herschel\/} mission that, following temperature drift removal, the SPIRE data is free enough from $1/f$ noise that most of the early SDP science goals could be achieved using a simple binning map-maker.  For the more sensitive science goals, we have developed an interim map-maker based on the algorithm of \cite{fixsen00}.  SHIM, the SPIRE-HerMES Iterative Mapper, exploits the redundant observations of each point on the sky with multiple detectors in cross-linked scan directions to determine both the sky signal and the weighting of the data samples for binning into map pixels while simultaneously producing an accurate error map.  Individual detector gains can also be determined, though in this instance the detector gains remain fixed.  SHIM works with SPIRE-only and SPIRE-PACS\footnote{PACS: Photodetector Array Camera and Spectrometer, \citealt{poglitch}} Parallel modes. The full algorithm is described in Section~\ref{sec:mapalg}. 

Following implementation, verification that the maps are free of artefacts from both the low level pipeline processing of time ordered data and of the mapping algorithm itself is required.   In order to characterize the effects of the timeline processing and the mapper itself, a detailed examination of the astrometry, noise properties, transfer function and other potential artefacts in the SHIM v1.0 maps is described in Section~\ref{systematics}.   Future versions of the SHIM pipeline are expected to build upon what is described here.  SHIM will be released as part of the HerMES initial data release (DR1), scheduled for the first quarter of 2011, via the HerMES website$^\dagger$ and the algorithm will ultimately be coded into java by the NASA {\sl Herschel\/} Science Center (NHSC) for inclusion in future versions of HIPE as a complement to the current naive mapper.

\section[]{Data Sets}
\label{sec:data}
The HerMES SDP fields are GOODS-N, Lockman-North and Lockman-SWIRE, FLS and the galaxy cluster Abell 2218.   GOODS-N, Lockman-North and Abell 2218 were all observed in SPIRE Large Map mode at the nominal scan speed of 30 arcsec/s.  The FLS field was observed in SPIRE-PACS Parallel mode, with a scan rate of 20 arcsec/s and the Lockman-SWIRE field was observed in SPIRE Large Map mode at the fast scan speed of 60 arcsec/s. Full details of the observed field sizes, scan strategies and depths are presented in \cite{Oliver}. Sections~\ref{sec:mapalg} and \ref{sec:tf} indicate that while the map depth, i.e. the number of detector samples per map pixel, is affected by the scan speed, the mapping algorithm and the expected/detected map systematics are independent of the observing mode among these fields.  The SHIM v1.0 maps of all five fields have been used by HerMES collaboration for several projects which rely on the statistical properties of the maps including \cite{nguyen}, \cite{glenn}, \cite{dowell2} and \cite{amblard}.

\section{Map Making Pipeline}
\label{analysis}

SHIM takes {\sc hipe\/}-processed timelines, calibrated in physical units, and injects them into an iterative mapper.  Both phases of this processing are described below.

\subsection{Timeline Processing}
\label{timestreams}

Raw telescope data were processed into calibrated timelines using {\sc hipe\/} version 3.0.1108 with the latest (at the time of processing) calibration files and modules  for removal of cosmic rays.  This processing also includes a corrections, provided by the SPIRE Instrument Control Center, of a 79\,ms time offset between the spacecraft and instrument clocks and a 2\,Hz correction to the approximately 300\,kHz clock frequencies of the spacecraft and instrument clocks\footnote{These correction are now part in the standard  {\sc hipe\/} pipeline.} to fix a pointing offset seen in early observations.  The details of the timeline processing are available in \cite{dowell} and in the SPIRE Observers Manual \citep{spireobs}.  In brief, once the raw data have been converted into detector voltages, several corrections to the timelines are made and the telescope pointing product is associated with each detector sample. The data are processed in the following order: first, detection and masking of cosmic ray glitches that affect all detectors (`frame hits') and then glitches that affect only individual detectors (`web hits'), in this case using the `simple' or `sigma-kappa' deglitcher which considers timeline data points that are $\kappa \times \sigma$ outliers to be glitches, where $\sigma$ is the standard deviation and $\kappa$ is chosen by the user.  In this analysis we use $\kappa = 5$.  Data affected by cosmic rays are flagged and are always masked at the map-making stage.  The signal timeline is interpolated over the affected samples to maintain a continuous timeline, but the interpolated data points are not included in the maps. The pipeline then corrects for the electrical filter response, converts signal to flux density and removes temperature drifts based on the FPA thermistors.  Finally, corrections are made for the bolometer time response before the signal timeline is merged with the telescope pointing product to produce sky coordinates for each detector at each sample.  While modules exist for the correction of electrical and optical crosstalk, but currently make no correction.  Calibrated timelines can then be fed into SHIM as described below.

\subsection{Mapping Algorithm}
\label{sec:mapalg}

Based on the algorithm described in  \cite{fixsen00}, SHIM uses an iterative baseline removal and calculation of detector
weights to produce 250, 350 and 500\,\micron\  maps with pixel sizes (6.0, 8.3 and 12.0 arcsec) chosen to be approximately $1/3$ the FWHM of the beam in the relevant band. The data are dealt with in scans, i.e. one pass across
the region of sky to be mapped. No timeline filtering is applied before the data are passed to the map-maker, except that the median
of each scan is subtracted, which increases  the efficiency with which the map-maker converges.
Given sky brightness, $M(x,y)$, the signal, $S_{dsj}$, for detector,
$d$, in scan, $s$, at time sample, $j$, is modeled as a sum of the sky, an offset and measurement noise, i.e.:
\begin{equation}
S_{dsj} = g_d \,M(x_{dsj}, y_{dsj}) + p^n_{ds} + N_{dsj},
\end{equation}
where $g_d$ is the detector gain, and $p^n_{ds}$ is an order-$n$
polynomial baseline offset for detector, $d$, and scan, $s$. $N_{dsj}$ is
the measurement noise, and $(x_{dsj}, y_{dsj})$ encodes the position
 of each detector on the sky. The initial sky map $M^0(x_{dsj}, y_{dsj})$ is a `naive' map in which the detector samples are simply binned into map pixels and the gains, $g_d$ are initialized to unity.  If the detector gains, $g_d$, are allowed to vary, the parameters $g_d$ and $p^n_{ds}$ are each fit iteratively, with each iteration, $i$,  consisting of two alternating steps. In the first step in each iteration, the $g_d$ are set to $g_d^{i-1}$ and are held fixed while the $n+1$ parameters, $p^{n}_{sd}$, for each detector, $d$, and scan, $s$, are varied simultaneously. Using a linear least squares minimization procedure the $p^{n}_{sd}$ are determined that minimize the variance of the timeline
residual, $R^i_{dsj}$,  based on the previous map, $M^{i-1}(x,y)$,
\begin{equation}
R^i_{dsj} = S_{dsj} - \left[ g^i_d \, M^{i-1}(x_{dsj},y_{dsj}) +
  p^{n,i}_{sd} \right].
\end{equation}
In the second step, for each detector, $d$, $g_d$ is varied to minimize $R_{dsj}$, while the $p^{n}_{sd}$ are fixed to $p^{n,i}_{sd}$.

A new estimate of the sky is then constructed by taking the weighted mean
of all samples that fall into each pixel:
\begin{equation}
M^i(x,y) = \frac{\displaystyle\sum_{dsj\,\in\,(x,y)} w^i_{ds} \, (S_{dsj} -
  p^{n,i}_{ds}) / g^i_d}{\displaystyle\sum_{dsj\,\in\,(x,y)} w^i_{ds}},
\end{equation}
where the weight $w^i_{ds}$ is the inverse variance of the timeline
residual,
\begin{equation}
w^i_{ds} = \left[\frac{1}{N} \sum^N_{j=1} \left(R^i_{dsj}\right)^2
  \right]^{-1},
\end{equation}
and $N$ is the number of samples in scan, $s$. Note that, while the weights are not free parameters, they are determined from the residuals as part of the iterative process.  A noise map is created
by propagating detector noise as estimated by the variance of the
residuals:
\begin{equation}
\sigma^i(x,y) = \left[ \sum_{dsj\,\in\,(x,y)} w^i_{ds} \right]^{-1/2}.
\end{equation}
The order $n$ of the polynomial baseline is chosen based on the physical size of the map where we use $n = 1$ for our smallest (9' x 9') cluster fields, $n=2$ for intermediate (30' x 30') deep fields and scale up to $n=3$ for the widest (2$^{\circ}$ x 2$^{\circ}$) fields. To improve stability of the
algorithm, the iteration on which to begin fitting each component (or whether to fit it at all) can
be specified. This includes the choice of whether to hold the $w_{ds}$ fixed.  The pipeline is typically run for several (5--10) iterations with the  $w_{ds}$ fixed to unity and allowing $w^i_{ds}$ to be determined by Eq. 4 in subsequent iterations.  In the maps presented here, the gains $g_d$ are fixed to unity. While the algorithm allows us to fit for the individual gains directly from the data this increase in complexity will be reserved for future versions of the maps.  The detector gains for the SPIRE bolometers have been derived using dedicated observations of calibration sources and the $g_d$ should not vary significantly from unity at this stage.  As a check, a single mapping run of SHIM on the GOODS-N field with $g_d$ allowed to vary produced $g_d$'s that differed from unity by less than a few percent.  While this may lead to a marginal improvement in the maps and will likely be implemented for future versions, the verification of this change is beyond the scope of the SDP phase.

The method described in Eq. 5 of estimating the noise map differs from the {\sc hipe\/} map-making pipeline, which uses the standard deviation of all samples that fall into each pixel.  Comparing the median noise in the error maps in both cases in the GOODS-N field, we find that they are not significantly different, indicating that the standard deviation provides a reasonable estimate of the overall noise level in the map. The direct propagation of detector noise used here should give a better measure of noise variations across the map and is more robust when the number of samples per pixel is low and the standard deviation becomes a poor estimate of the statistical error in the measurement.  While, in general, only a small fraction of pixels ($< 0.1 \%$) have less than 5 samples, in fast scan mode at 250\,\micron, where 7 dead or slow bolometers are masked, regions of the map that have had full scans masked due to temperature drift artefacts have pixels with as low as 2-3 samples.  Splitting the data in half for the jackknife tests described in the following sections then leads to occasional pixels with only a single sample, for which the standard deviation can not provide an error estimate.  It should be noted that the FLS field, observed in SPIRE-PACS Parallel mode scanning at 20 arcsec/s in only a single repeat, has fewer low sample pixels than the two repeat observation of Lockman-SWIRE field observed in SPIRE Large Map mode scanning at 60 arcsec/s just as one would expect given three times more samples per pixel scanning at 20 vs. 60 arcsec/s.  

\begin{figure}
\centering
\includegraphics[width=\linewidth]{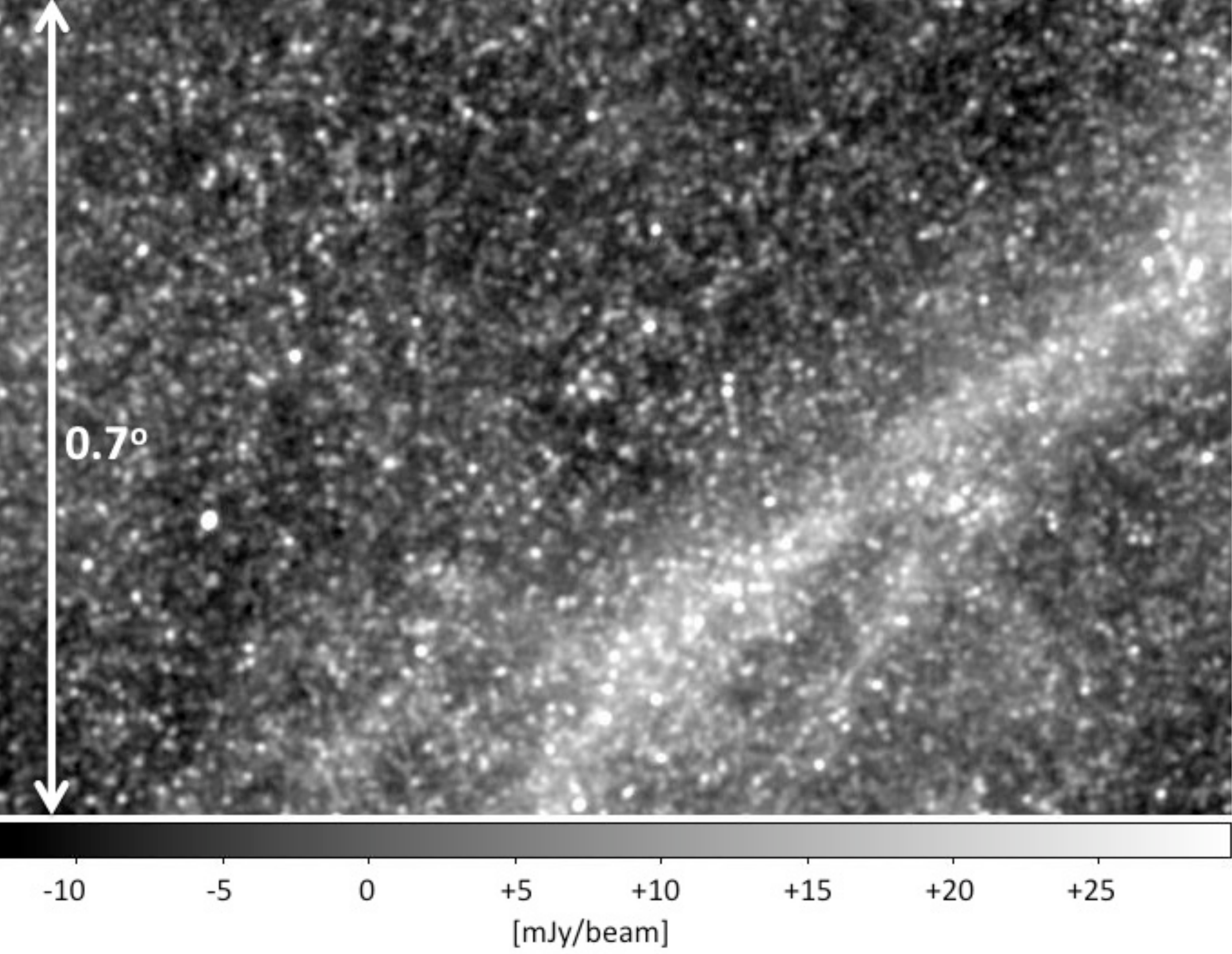}
\caption{A 0.7 deg x  1.1 deg sample of the FLS field displaying both the confused SPIRE point source background and a bright strand of Galactic cirrus.}
\label{fig:images}
\end{figure}

 High quality SHIM maps have been produced for all HerMES SDP fields.  A  0.7 deg x 1.1 deg portion of the 250\,\micron\ FLS field is shown in Fig.~\ref{fig:images}. This particular field is chosen to deliberately illustrate the range of structures on which SHIM is designed to operate as it has a strong band of Galactic cirrus. Cirrus structure like this is not typical of the HerMES fields and, in fact, the FLS field has been excluded from statistical studies for HerMES due to the cirrus contamination.  While all of the HerMES maps are visually flat and artefact free, a program of verification, described in the sections below, was performed to determine on what spatial and point source flux scales these maps are reliable for use in scientific analysis.

\section{Systematics Control}
\label{systematics}

We describe below the various tests we apply to test for systematic effects in the SHIM maps.  This includes a correction of the absolute astrometry (Section~\ref{sec:astrom}), detailed analysis of the noise properties of the maps based on jackknife (difference) maps (Section~\ref{jacks} \& Section~\ref{powerspec}), measurement of the map transfer function in both deep and wide maps (Section~\ref{sec:tf}) based on simulated beam-scale and large-scale maps and finally a characterization and bound on the effect of a small, correlated relative pointing error seen in the deepest maps (Section~\ref{sec:pointingdrift}).  

\subsection{Absolute Astrometry}
\label{sec:astrom}

\begin{figure}
\centering
\includegraphics[width=\linewidth]{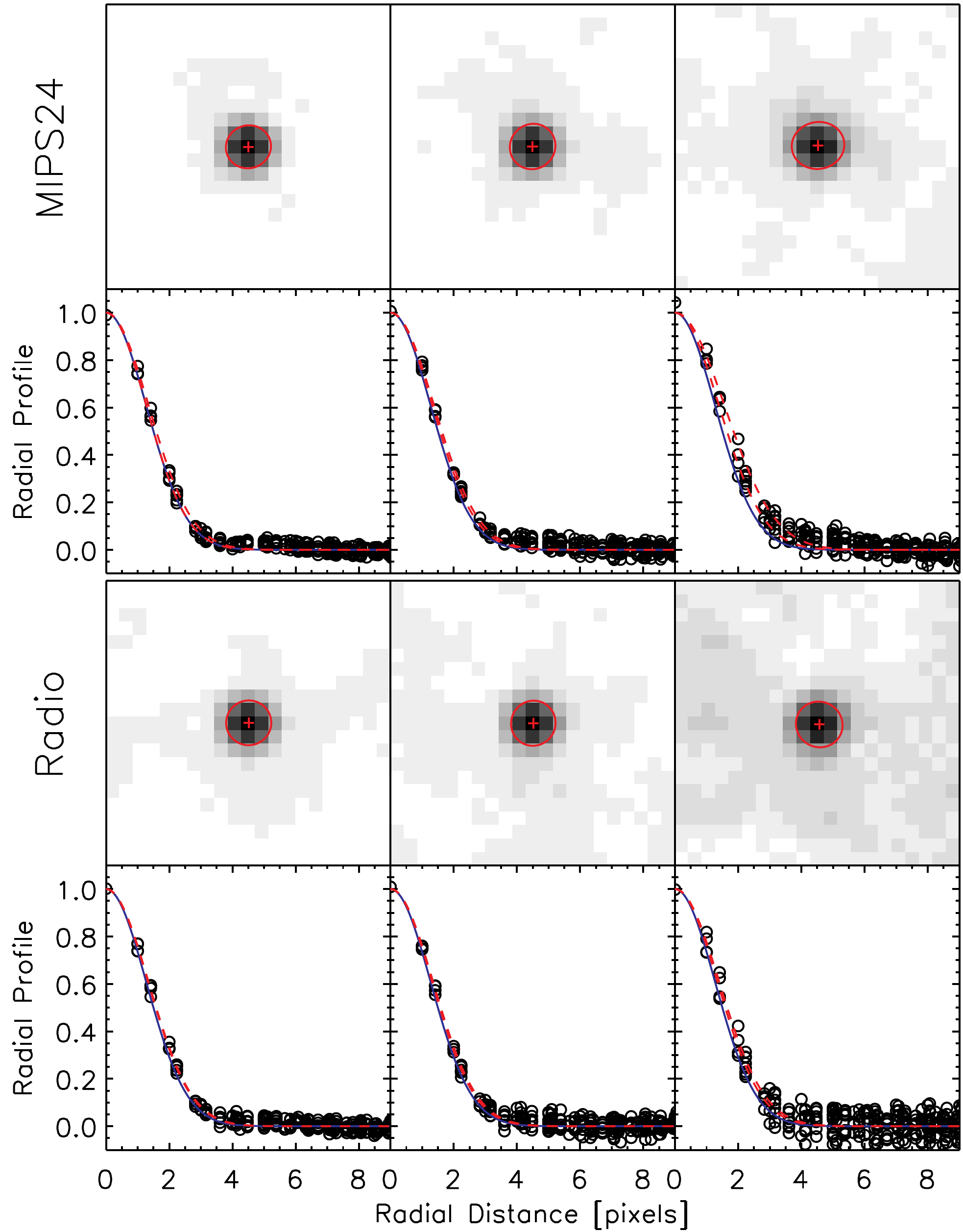}
\caption{The GOODS-N SHIM maps stacked at the location of
  24\,\micron\ (upper panels) and radio sources (lower
  panels). Signal-to-noise ratio images and radial profiles at 
  250, 350 and 500\,\micron, from right to left. The FWHM ellipses
  corresponding to the 2-d Gaussian fits are overlaid on the
  images. The major and minor radial profiles (red dashed lines) and
  nominal beam profiles (blue solid lines) are also shown. Pixel sizes are
  6, 8.33 and 12\,arcsec at 250, 350 and 500\,\micron, respectively. Peak positions are used to determine pointing offsets and the good agreement with the nominal beam shapes indicates the result is not heavily biased due to clustered, confused sources in the stack.}
\label{fig:stacks}
\end{figure}

The absolute astrometry of the data is measured by stacking the SPIRE maps at the positions of MIPS 24\,\micron\ and radio sources.  24\,\micron\ source catalogs come from the {\em Spitzer\/} Multi-Band Imaging Photometer
(MIPS)  (\cite{werner}, \cite{rieke}).  These include the publicly available GOODS catalog and catalogs from \cite{shupe} (Lockman-SWIRE) and \cite{fadda} (FLS).  Comparisons were also made to radio source catalogs in all fields including \cite{owen} (Lockman-North), though the corrections to the astrometry are based on the 24\,\micron\  catalogs. Fig.~\ref{fig:stacks} shows the results of the stacks along with the centroids and FWHM ellipses resulting from a 2d Gaussian fit to the stack where a 2d fit is chosen due to the known, small mean ellipticities of the SPIRE beams of (7\,\%, 12\,\%, 9\,\%) \citep{griffin}.  The Figure also shows that while the 24\,\micron\ stacks are slightly broader than the nominal beam sizes, particularly at 500\,\micron, the radio stacks are well described by the nominal beams sizes of 18.1, 25.2 and 36.6\,arcsec \citep{swinyard}.  

The absolute pointing accuracy of the MIPS catalog sources is better than one arcsec, as determined from matching 2MASS sources in the SWIRE MIPS fields.  Radio stacks are consistent (within 0.5 arcsec) with the MIPS stacks in Lockman-SWIRE and the other HerMES fields and an individual overall correction to the absolute astrometry, on the order of a few arcseconds, was made to each SPIRE map based on the 24\,\micron\ stacks.

In all cases, the number of sources used in the MIPS stacks far exceeds the number of sources for the radio (by factors of 10, 4, 30, 8, and 100, for the five fields). We correct the v1.0 maps based on  the higher-SNR MIPS stacks (particularly for Abell 2218, where there are only 10 radio sources in the catalog). The offsets are not consistent from field-to-field since the pointing offset is known to be fixed in telescope coordinates and not celestial coordinates, so depending on the position angle of a given field, the offsets will be different in celestial coordinates.  These offsets have now been corrected in the {\sc hipe\/} pipeline and should no longer be necessary at the map-making stage.

\subsection{Jackknife Tests}
\label{jacks}

\begin{figure}
\centering
\includegraphics[width=\linewidth]{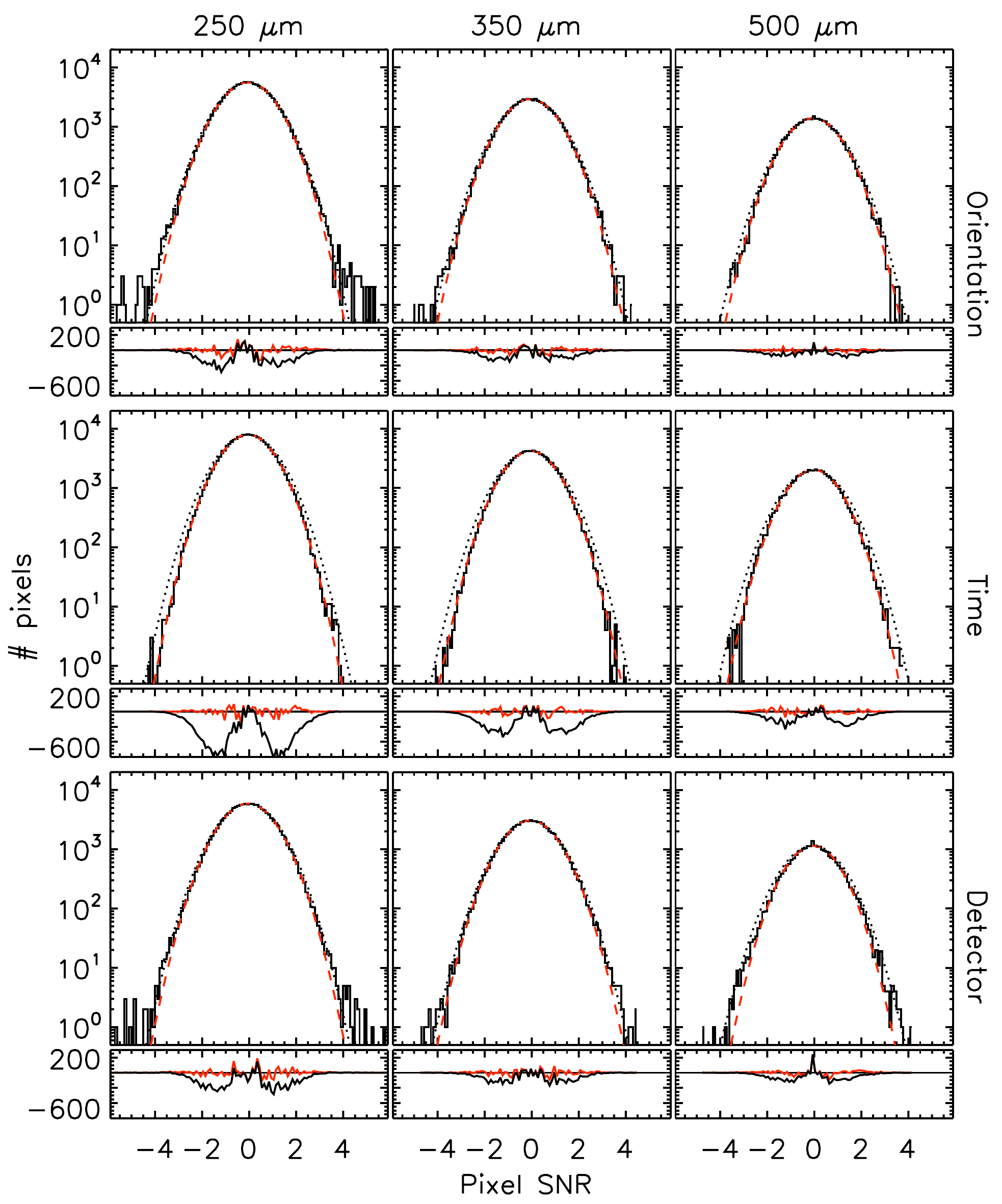}
\caption{Distribution of map signal-to-noise ratio measured from
  jackknife maps in the GOODS-N field. Three different cuts of the
  data are shown: (i) `orientation' compares scans across the field
  at one position angle compared to the orthogonal direction; (ii)
  `time' compares the first and second halves of the data, split by
  observing time; and (iii) `detector' compares data collected from
  one half of the detectors in each band (alternating detectors are chosen)
  compared to the other half. The best-fit Gaussians are shown as
  dashed red curves and the expected $\sigma$ = 1 Gaussians are shown as dotted lines. Below each histogram is a plot of the histogram residuals with the best-fit residuals in red and the residuals from the $\sigma$ = 1 histograms in black.}
\label{fig:jackknife}
\end{figure}

We divide the data into two halves in various ways to examine the
noise properties. Figure~\ref{fig:jackknife} shows the distribution of
signal-to-noise ratio from the three types of jackknife difference
maps, `orientation', `time' and `detector', for each
band. `Orientation' splits the data by the position angle of the
scans across the field into (nearly) orthogonal sets. `Time' splits the data
into halves based on observation time. `Detector' divides the focal
plane at each band into two sets of detectors, each of which approximately covers the full spatial extent of the
focal plane. The sub-maps are made with the mapping parameters
($p^n_{ds}$, $g_d$ and $w_{ds}$) determined from the full data
set. For each band and type of jackknife, the sub-maps are subtracted
and the distribution of pixel SNRs are examined. We see from the overplotted best-fit Gaussians and the residuals below each histogram that they are
very well described by Gaussian curves, as one would hope if the map noise is uncorrelated.  The 1\,$\sigma$ Gaussian widths of these histograms are: Orientation (0.95, 0.95, 0.94), Time (0.91, 0.91, 0.90), Detector (0.95, 0.95, 0.89) at 250, 350 and 500\,\micron\ respectively. Assuming that the jackknife difference maps give an accurate measurement of the map noise, the slight deviation from the expected zero-mean, $\sigma$ = 1 Gaussians (particularly noticeable in the the black residual curves) indicates that the noise maps are biased high by 5 to 10 percent in all cases. For this reason, we recommend using the jackknife difference maps to accurately assess the noise in the maps.

\begin{figure}
\centering
\includegraphics[width=\linewidth]{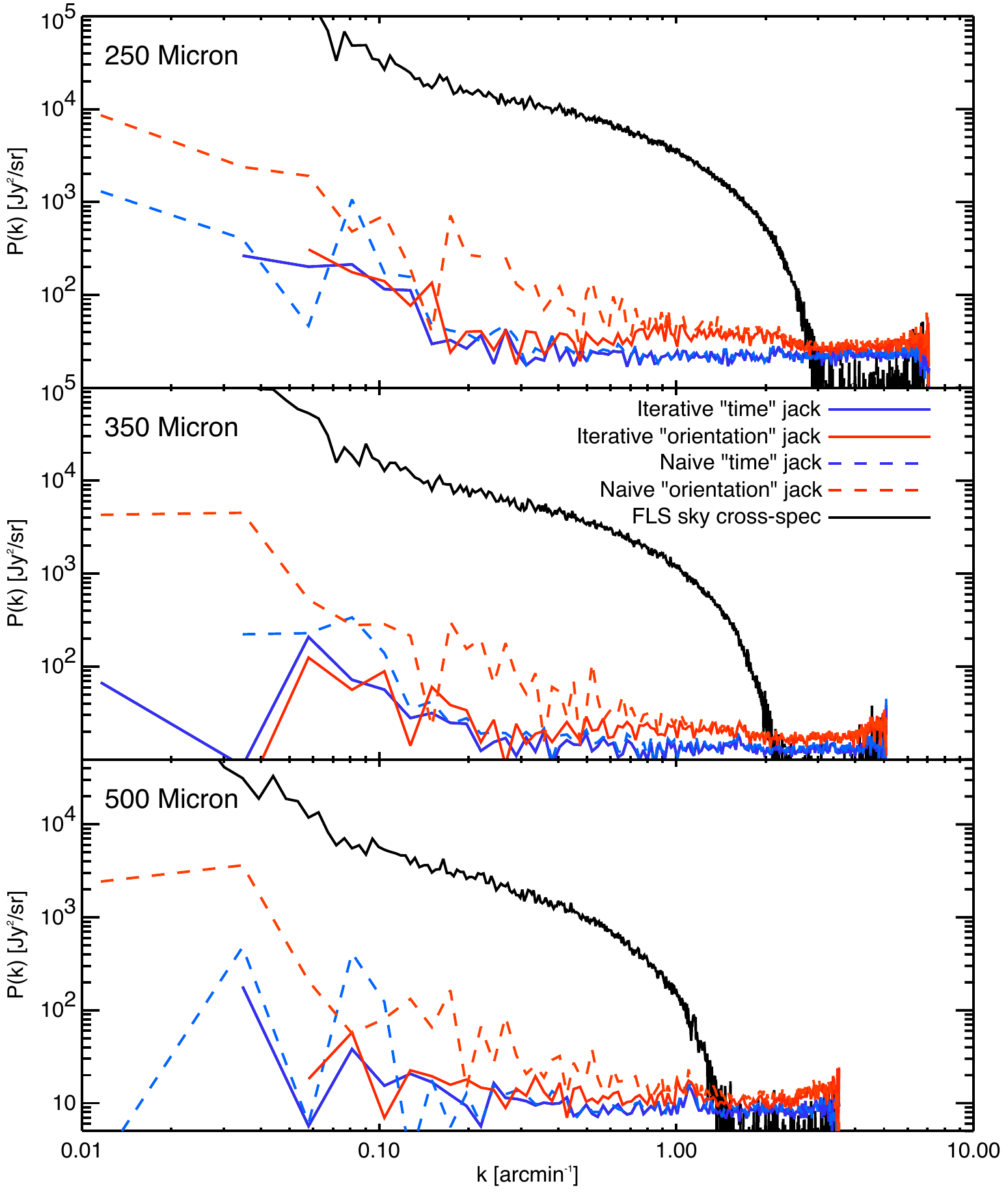}
\caption{ Auto power spectra of the difference jackknife maps of the GOODS-N field, separating the scans both by time (blue lines) and 
by scan orientation (red lines),  for each band (250 to 500\,\micron\ from top to bottom) 
with both naive  (dashed lines) and SHIM maps (solid lines). These show that even in maps with only a single scan orientation, SHIM removes low frequency noise on scales below 1 degree.  For comparison, the solid black lines
represent the \it cross\rm-spectrum of the astrophysical signal as measured in the FLS field.}
\label{fig:powerspec}
\end{figure}

\subsection{Noise Power Spectra}
\label{powerspec}

Auto power spectra of both the `time' and `orientation' jackknife difference maps were produced from the GOODS-N maps after accounting for partial sky coverage. The power spectra are shown in Figure~\ref{fig:powerspec}.  These power spectra were produced for the map built with a simple, `naive', projection algorithm  as well as with the iterative SHIM algorithm.  In the naive maps, the `orientation' jackknife contains more residual noise than the `time' jackknife since the two different scan orientations in the `time' map allow better removal of the $1/f$ noise. In both cases, however, the iterative map-maker allows us to mitigate this effect on scales larger than the instrument beam. The `time' jackknife spectra are almost identical whether they are computed with the `naive' or
iterative map-maker due to the large number of scans and the two different scan orientations. For comparison, the \it cross\rm-spectrum of the astrophysical signal in the FLS field is also shown (in black) in Figure~\ref{fig:powerspec}.  This spectrum is obtained by correlating two maps of the FLS field using 
two independent halves of the FLS data, split in time.  Since the noise between the two halves of the data is uncorrelated, the cross-spectrum shows the astrophysical power with the instrument noise removed.   The FLS field contains more Galactic cirrus signal than the GOODS-N field,
but the extragalactic background, dominant at $k$ larger than 0.1 arcmin$^{-1}$, should be statistically equivalent. The FLS cross-spectrum is therefore a good approximation of the astrophysical power in GOODS-N and
Fig.~\ref{fig:powerspec} shows that the astrophysical power is dominated by the instrument noise on scales smaller than the beam FWHM in each band."

The instrumental noise power spectrum amplitude, compared in the same region, is 10 to 100 times lower than the astrophysical signal on scales larger than  the beam.
On scales larger than 10 arcminutes, there does appear to be some additional power left compared to a pure white noise spectrum. This extra power may originate from residual temperature drifts which are currently only modeled
by 2nd order polynomials.  In order to trace and bound the effect of this excess power on specific scales, in the next sub-section, we describe the use of simulations  to determine point source fidelity and  to determine the SHIM transfer function (Section~\ref{sec:tf}). We investigate a small but correlated relative pointing error (Section~\ref{sec:pointingdrift}) and  we also comment in Section~\ref{sec:concl} on low level residual effects due to timeline processing of the SPIRE data that may remain at this early stage of the {\sl Herschel\/} mission.

\subsection{Map Transfer Function}
\label{sec:tf}

The map transfer function for the SHIM SDP maps, i.e. the correction in Fourier space for the effects of the mapper based on the ratio of the Fourier transforms of the  simulated maps and the input skies for that simulation, has been measured using simulated datasets including beam-scale-only power, and separately, including Gaussian fluctuations on all scales.   A deviation of the transfer function from unity indicates power either added or suppressed in the maps by the mapper itself.  Where deviations from unity are small ($\lesssim \pm$ 10-15\,\%) measurements of angular power in a map must be corrected with this transfer function to recover the true power on the sky.  Larger deviations indicate that the maps are no longer reliable at the corresponding angular scale.

\begin{figure}
\centering
\includegraphics[width=\linewidth]{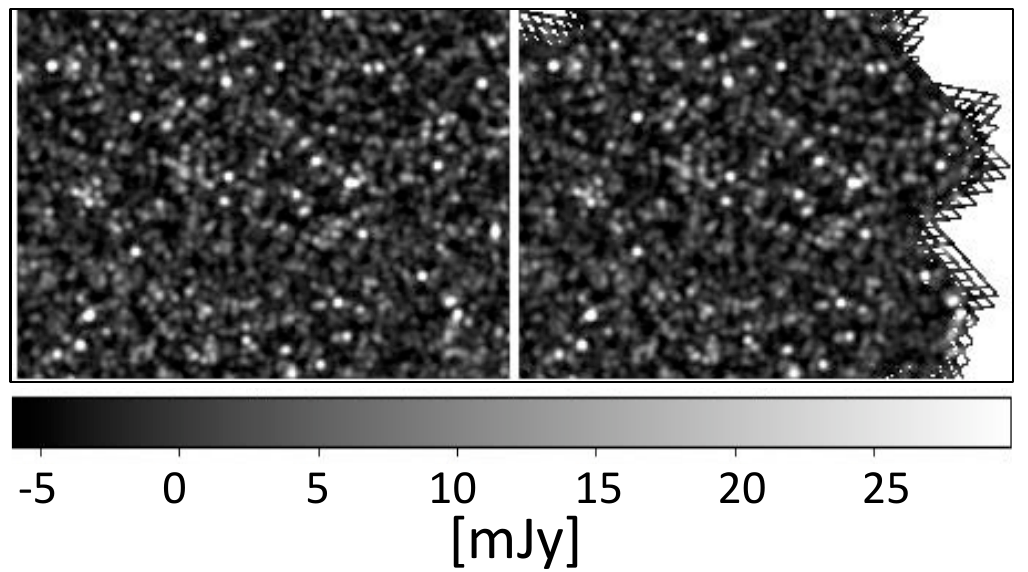}
\caption{One of the 50 simulated noiseless realizations of the BLAST sky are shown in both input (left) and SHIM output (right) at 250\,\micron. The input map is simply a simulated point source field convolved with a Gaussian beam of FHWM equal to that of the SPIRE beam at 250\,\micron.  The output map shows the characteristic scan pattern of the SPIRE bolometer arrays as the input map has been resampled into timelines with the exact scan pattern of the real observed map.  The resampled timelines are then run through the SHIM to produce the map on the right. }
\label{fig:tfpic}
\end{figure}

\begin{figure}
\centering
\includegraphics[width=\linewidth]{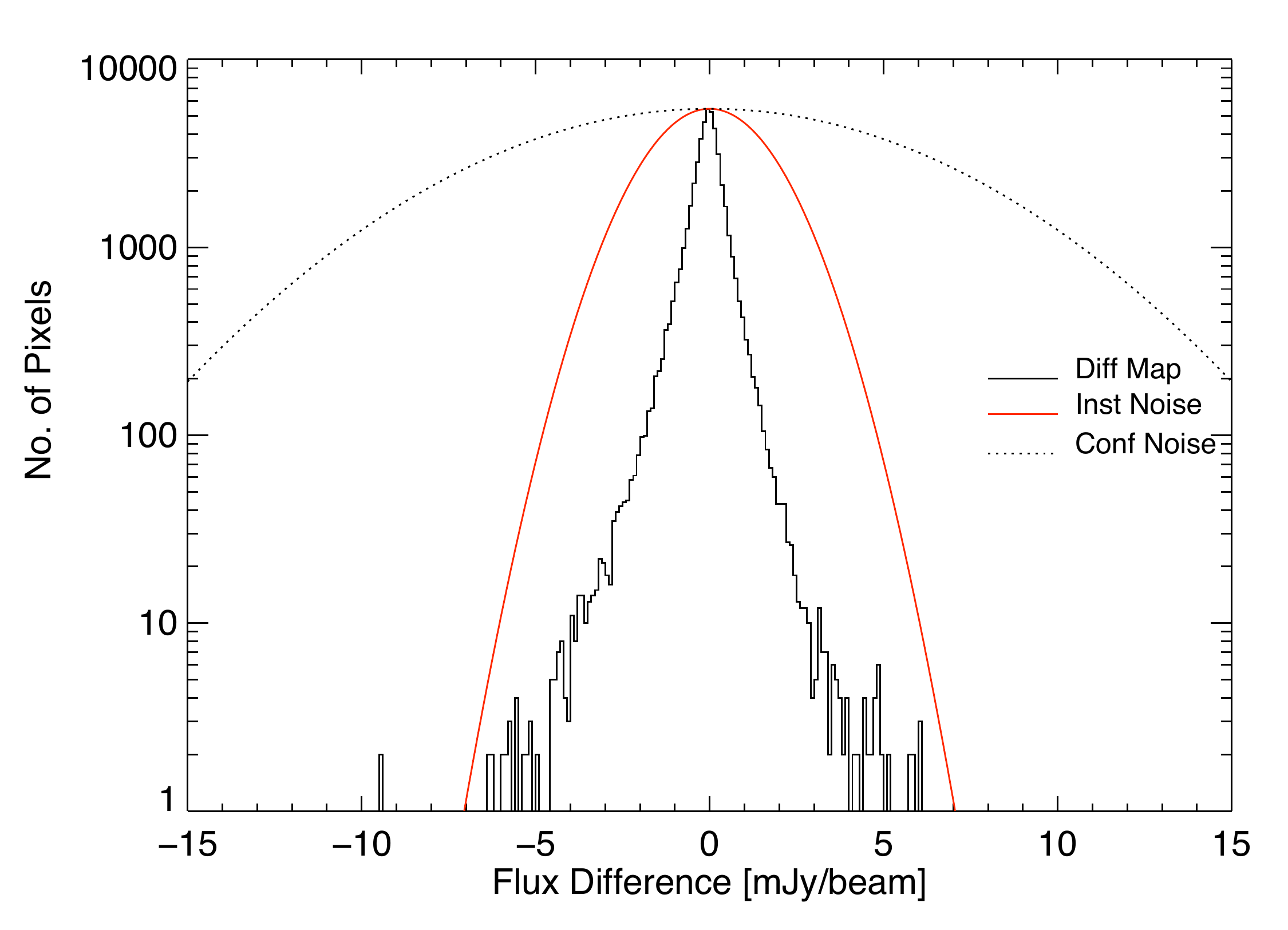}
\caption{Pixel histogram of the average of the noiseless, simulated input maps minus the resulting SHIM output  maps.   Overplotted in red is a Gaussian with $\sigma = 1.7$\,mJy beam$^{-1}$, the width set by the instrument noise in the deep GOODS-N map.  Also shown (dotted) is a Gaussian showing the confusion noise with $\sigma = 5.8$\,mJy beam$^{-1}$.  The interquartile range of the histogram is 0.4 mJy\,beam$^{-1}$. }
\label{fig:tf}
\end{figure}

Using the best fit galaxy counts model from BLAST \citep[the Balloon Borne Large Area Submillimeter Telescope,][]{pat09}, fifty noiseless realizations of the 250, 350 and 500\,\micron\ sky were created.  While improved counts from \cite{Oliver} at bright fluxes and \cite{glenn} at faint fluxes now supersede the BLAST counts, these counts provide a reasonable model of the SPIRE sky and were readily available for SPIRE SDP work.  Simulated skies were resampled into timelines and fed, in the same way the real  data, into the SHIM iterative mapping pipeline.  Input and output maps for one of the 50 realizations of the simulated sky are shown in Figure~\ref{fig:tfpic}.  Figure~\ref{fig:tf} shows the histogram of the average of 50 difference maps made by subtracting the simulated input point source sky (left panel of Figure~\ref{fig:tfpic}) and the SHIM map (right panel of Figure~\ref{fig:tfpic}).   A Gaussian with a width corresponding to the (1$\sigma$) instrument noise in the deep GOODS-N map ($\sigma_{inst}$ = 1.7 mJy\,beam$^{-1}$) is shown overplotted in red.  Also shown as a dotted line is a Gaussian with width corresponding to the  (1$\sigma$) confusion noise ($\sigma_{conf}$ = 5.8 mJy\,beam$^{-1}$, \cite{nguyen}).  The interquartile range of the histogram is 0.4\,mJy\,beam$^{-1}$.  Since this includes any errors in rendering the point sources and any other signal introduced by the mapper, this range gives an indication of an upper limit on the spread in point source errors. For a source at the SPIRE 250\,\micron\ confusion limit (3$\sigma_{conf}$ $\approx$ 18\,mJy), this corresponds to a 2.2\,\% error.  The current SPIRE flux calibration accuracy is quoted as $\pm 15\,\%$ \citep{swinyard}.  Therefore, errors due to the map-maker at the 2\,\% level are negligible in relation to the current calibration uncertainties.

\begin{figure}
\centering
\includegraphics[width=\linewidth]{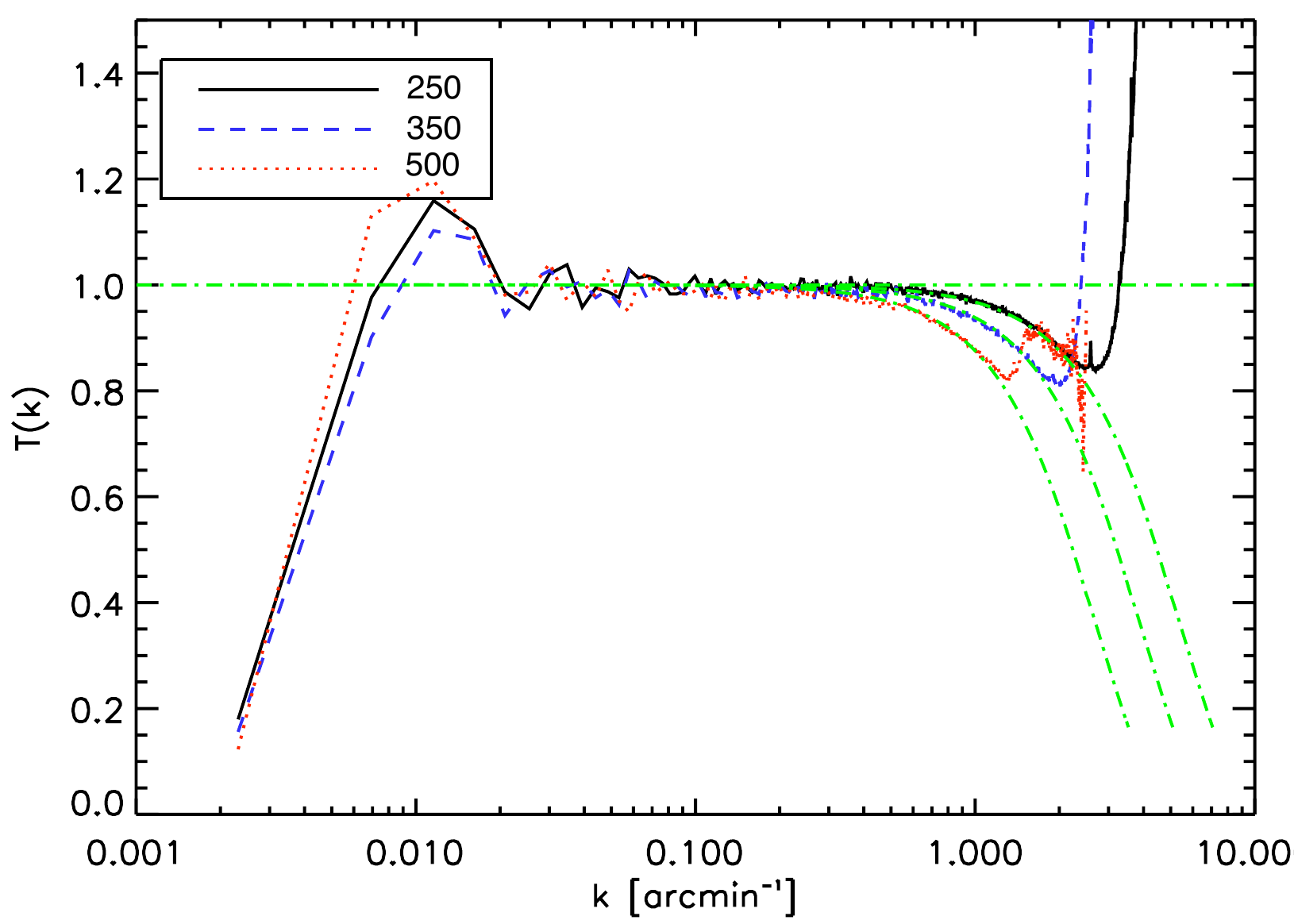}
\includegraphics[width=\linewidth]{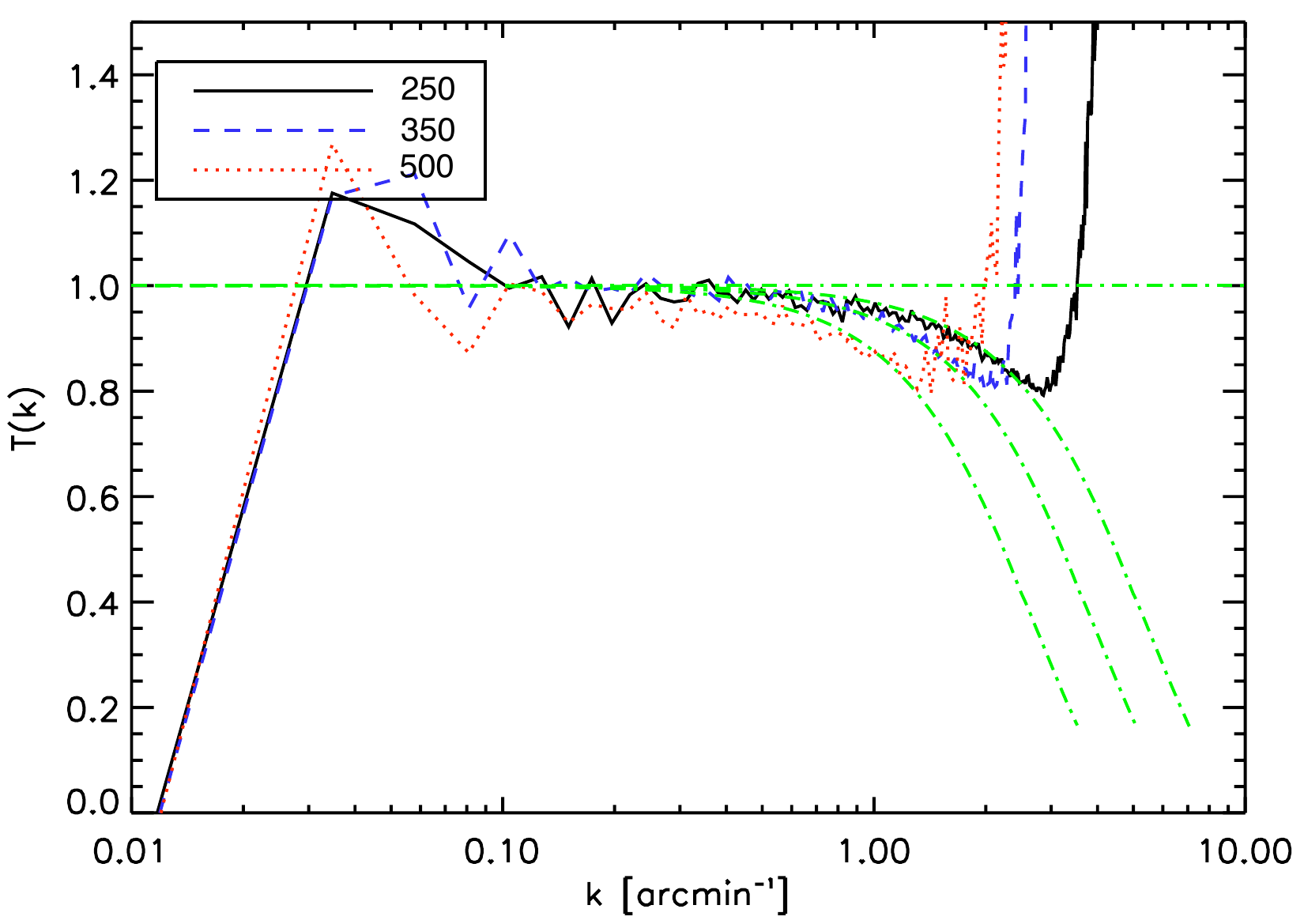}
\caption{Transfer function, $T(k)$, of the simulated large-scale structures in the wide Lockman-SWIRE (top) map, observed at 60 arcsec/s and 
the deep GOODS-N (bottom) map, observed at the nominal scan speed of 30 arcsec/s. The transfer functions are well described by the pixel window functions
(green dot-dashed lines) from 0.02 arcmin$^{-1}$ to about 1 to 3 arcmin$^{-1}$. The transfer functions show about 20 per cent extra power on the scale of half size of the map (0.01 arcmin$^{-1}$ (top) and 0.07 arcmin$^{-1}$ (bottom)).}
\label{fig:tf_lss}
\end{figure}

Testing for mapping effects on large-scale structures has been performed using 100 simulated realizations of beam-convolved Gaussian sky fluctuations, as presented in Figure~\ref{fig:tf_lss}. Shown are the auto-spectrum transfer functions at all three SPIRE wavelengths for two different fields, GOODS-N and Lockman-SWIRE, with respective map widths of 30 arcmin and 4 deg and scan speeds of 30 and 60 arcsec/s. 
As in the above analysis, simulated maps are resampled into timelines which are then fed to SHIM.  To interpret these transfer functions, the dot-dashed green lines show the expected transfer function due to the pixel size effect, 
described by the function $(\mathrm{sinc}(k_\mathrm{x}\theta_{\rm pix}/2)\times\mathrm{sinc}(k_\mathrm{y}\theta_{\rm pix}/2))^2$, a necessary consequence of using a finite pixel size.
This effect is detectable in the transfer function since the input simulations have a much finer (2 arcsec) resolution.  Apart from this pixelization effect, which must be corrected for in measurements of the angular power in the SPIRE maps, the mapper transfer function is approximately unity from scales from
around quarter of the size of the survey (roughly 0.02 and 0.1 arcmin$^{-1}$ for Lockman-SWIRE and GOODS-N respectively)
 to the scale of the FWHM for the relevant band (1.5 to 3 arcmin$^{-1}$). 
 
 At angular scales just above the beam FWHM in each band, this transfer function fails.  While primary beam effects largely cancel in this ratio, since both the input simulated maps and the output SHIM maps have the same beam, the effect of the mapper slightly alters the beam widths.  As shown above, this only amounts to a small error in the total point source fluxes, but a tiny amount of excess power at these scales, where the input simulations have zero power, creates an explosion of the transfer function.  As demonstrated in \cite{amblard}, this transfer function correction is no longer viable for $k \gtrsim$ 1, 2 \& 3 arcmin$^{-1}$ at 250, 350 and 500\,\micron, respectively.  \cite{glenn} detail the separate set of simulations  required for their study of the beam scale power in the HerMES maps.
 The iterative mapper seems to create 20\,\% extra power at about half  the size of the map (here 0.01 and 0.05 arcmin$^{-1}$).  This effect is dependent on the scan length, and could be an effect of the choice of polynomial order.  Current measurements on these scales should correct for this effect and this mapping artefact will be addressed in future data releases.   SHIM is, however, very clean at most angular scales under the realistic conditions for which we tested it and we note that the map transfer function does not appear to depend on the scan speed.
 
\subsection{Relative Pointing}
\label{sec:pointingdrift}

\begin{figure}
\centering
\includegraphics[width=\linewidth]{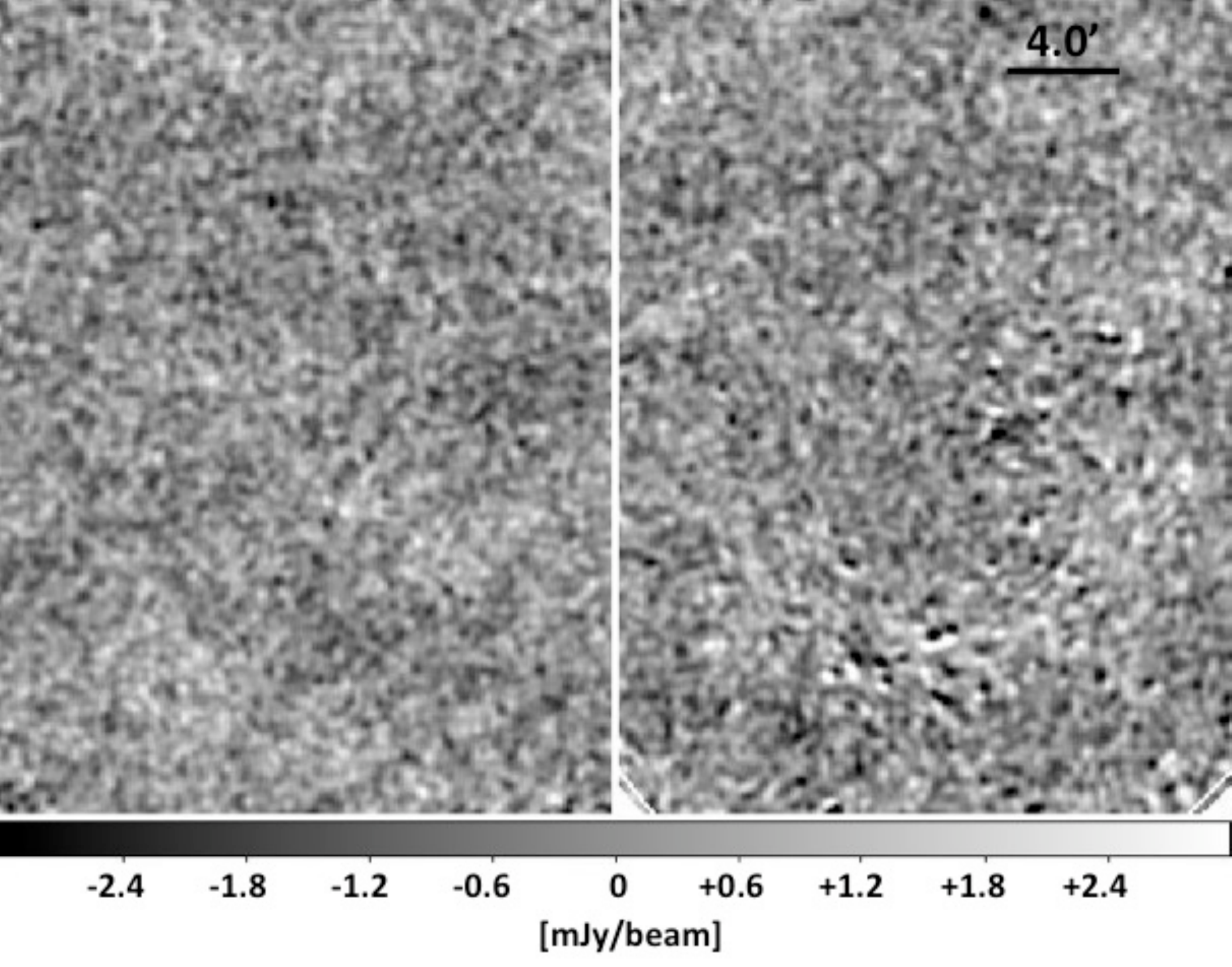}
\caption{Time (left) and Orientation (right) jackknife maps of the GOODS-N field smoothed with an 18 arcsec FWHM Gaussian kernal. The right panel shows a small residual correlated pointing error enhanced enhanced, for visual purposes, by the smoothing function.}
\label{fig:warp}
\end{figure}

In the deepest maps, we see a small but correlated pointing error.  This is a correlated shift of the point source positions, generally on the order of 1 arcsec, with maximum errors as large as 3 arcsec, the detectability of which depends on the depth of the observation.  Figure~\ref{fig:warp} shows a portion of the GOODS-N jackknife maps.  The left panel is the `time' jackknife in which the residual noise appears white.  The right panel shows the `orientation' jackknife in which we see small, correlated residuals rising above the noise.  As seen in the nearly Gaussian histograms in Figure~\ref{fig:stacks}, the effect on the overall map statistics is small.  
The artefact has not been fully characterized at this early stage of the {\sl Herschel\/} mission, however,  and we urge caution for those performing detailed statistical analyses of the SPIRE maps, whether they are the maps described in this work or maps produced independently by HIPE or another, custom map-maker.  

\begin{figure}
\centering
\includegraphics[width=\linewidth]{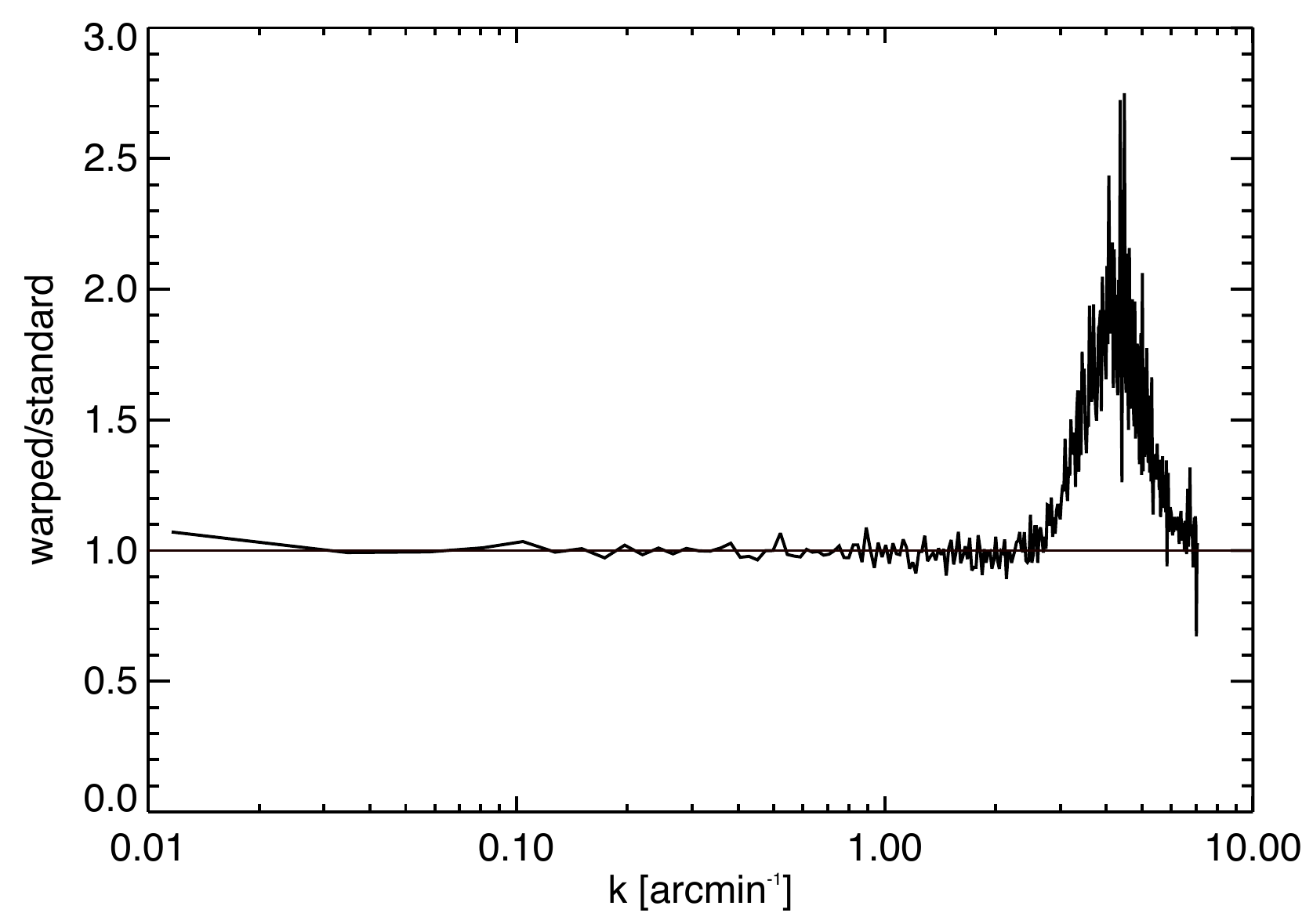}
\caption{The ratio of the power in a 250\,\micron\ simulated map with a highly exaggerated pointing error (warping) model to the power in a map from the same simulation without pointing error.  The warping model, meant to bound the effect of the pointing error seen in Figure~\ref{fig:warp}, has pointing errors  5 times larger than the largest errors seen in the true maps and cover the entire field.  In reality, only some regions of the map are affected and the amplitude of the shift is less than 3 arcsec.  The excess near 4 arcmin$^{-1}$, however, indicates the scale on which this pointing error may be important for statistical studies.}
\label{fig:warpmodel}
\end{figure}

In order to determine the angular scales on which this issue is important and to bound a worst-case limit, we developed a simple field `warping' model which creates correlated pointing errors throughout a simulated map.  The warping effect is visually reminiscent of a much more pronounced warping seen in early data due to the timing offset between the SPIRE and spacecraft clocks discussed in Section~\ref{sec:data}. This issue of clock drift is now well understood and a permanent solution has been implemented in the HIPE pipeline.  However, while the underlying cause of the remaining warp is likely due to another effect and is so far not  well understood, a model for the field warping that produces an error with similar features to the observed error can be  created easily by re-introducing a continuous clock drift into a simulated observation between the detector positions and the associated pointing.   The modeled drift was chosen to create an amplitude of the errors five times larger than the largest shifts seen in the sky maps.  Figure~\ref{fig:warpmodel} shows the ratio of the power in the 250\,\micron\ simulated map with warping to the power in a map from the same simulation without warping.  The excess power seen near 4 arcmin$^{-1}$ indicates that this effect will be seen most strongly in the map power on angular scales just below the beam diameter.   In real data, the effect in the science maps is small, as seen in the  
jackknife power spectra in Fig.~\ref{fig:powerspec}.  While the white noise level is slightly higher in the `orientation' jackknife maps, there is no significant bump near 4 arcmin$^{-1}$ as one might expect given the
simulated $P(k)$ in Figure~\ref{fig:warpmodel}.  This indicates that this is a minor issue, but it is one that in not yet corrected for in either HIPE or SHIM and should be noted. The effect of this shift has been simulated and bounded by those HerMES SDP papers that are sensitive to these kinds of issues, e.g \cite{glenn} and  \cite{amblard}.

\section{Conclusions}
\label{sec:concl}

The HerMES team has designed and implemented a map-making pipeline, called SHIM, for the production of maps from {\sc hipe\/}-processed, SPIRE data observed in SPIRE Large Map or SPIRE-PACS Parallel modes. SHIM reproduces both point source and diffuse extended emission with the highest possible fidelity for the most sensitive statistical analyses in maps with pixel sizes of 6.0, 8.3 and 12.0 arcsec in the SPIRE bands centered at 250, 350 and 500 \,\micron .   A rigorous program of investigation into the noise properties of the SHIM maps, their relative and absolute astrometry, the map transfer function and low level residual effects, shows that  the HerMES SDP maps are reliable on scales from the beam diameter up to at least 1 degree. Transfer function corrections to both point sources and large-scale structure are at the level of $<5\,\%$.  Since these effects are smaller than the current overall flux calibration error of $\pm$ 15\,\%, we have determined that the v1.0 SHIM maps are reliable for large-scale structure studies of the FIR/submm Universe using the bulk statistical properties of the SPIRE maps on angular scales from beam size up to about half the map area.  Residual mapping errors at the few percent level, such as the pointing drift artefact,  and their effect on sensitive SDP science have been flagged here.  The HerMES statistical analyses most at risk of being affected by such residuals,  \cite{glenn} and \cite{amblard}, have bounded the effects on their science results using simulations as described in those papers. The SHIM maps exhibit about 20 per cent excess power at scales larger than one degree and analysis requiring these scales should address the potential effect of this excess power on their results.  Timeline processing modules in {\sc hipe\/} that deal with temperature drift removal and cosmic ray removal are still under development.  Currently, the temperature drift removed from the detector signal is based on 2nd order polynomial models of the thermistor signal.  Residual temperature drift left in the detector signal, as well as potential  glitches missed by the current glitch removal modules could conceivably be leaving artefacts in the maps which have been bounded here and are expected to be continually improved. As in the case of the pointing error, those science efforts aimed at pushing the limits of the SPIRE data should characterize and bound the relevant effects on their science results based on the SDP processing of the SPIRE data. As the {\sl Herschel\/} mission moves from SDP into Routine Observations, we anticipate improvement in both the {\sc hipe\/} timeline processing and map making.  Increasing sophistication in the SHIM implementation will likely lead us toward to an optimum map-maker that, along with improved timeline processing will extend the reliability of the HerMES maps well beyond the limits described here.  The software discussed here is scheduled for release by HerMES along with the DR1 and is now being coded by the NHSC for inclusion in HIPE.

\section*{Acknowledgments}
SPIRE has been developed by a consortium of institutes led by Cardiff Univ. (UK) and including Univ. Lethbridge (Canada); NAOC (China);  CEA, LAM (France); IFSI, Univ. Padua (Italy); IAC (Spain); Stockholm  Observatory (Sweden); Imperial College London, RAL, UCL-MSSL, UKATC, Univ. Sussex (UK); Caltech, JPL, NHSC, Univ. Colorado (USA). This development has  been supported by national funding agencies: CSA (Canada); NAOC (China);  CEA, CNES, CNRS (France); ASI (Italy); MCINN (Spain); SNSB (Sweden); STFC  (UK); and NASA (USA). The authors thank Dave Clements and Mattia Vaccari for useful comments.

\bsp

\label{lastpage}

\end{document}